\documentclass[12pt,preprint]{aastex}
\usepackage{graphicx}
\usepackage{epsfig}

\def\Msol{\thinspace\hbox{$\hbox{M}_{\odot}$}}

\def\a4{\hsize 17.0cm \vsize 25.cm}
\newcommand{\der}[2]  { \frac{{\rm d}#1}{{\rm d}#2} }

\shorttitle{Spherically-symmetric Accretion onto a BH inside YSC}
\shortauthors{Silich et al.}

\begin{document}

\title{Spherically-symmetric Accretion onto a Black Hole at the Center of a
       Young Stellar Cluster}

\author{
Sergiy Silich, Guillermo Tenorio-Tagle
\and 
Filiberto Hueyotl-Zahuantitla
\affil{Instituto Nacional de Astrof\'\i sica Optica y
Electr\'onica, AP 51, 72000 Puebla, M\'exico; silich@inaoep.mx}}

\begin{abstract}
Here we present a self-consistent, bimodal stationary solution for 
spherically symmetric flows driven by young massive stellar clusters with
a central supermassive black hole (SMBH). We demonstrate that the 
hydrodynamic regime of the flow depends on the location of the cluster in 
the 3D (star cluster mechanical luminosity - BH mass - star cluster radius)
parameter space. We show that a threshold mechanical luminosity ($L_{crit}$)
separates
clusters which evolve in the BH dominated regime frome those whose internal
structure is strongly affected by the radiative cooling. In the first case
(below the threshold energy) gravity of the BH separates the flow into two 
distinct zones: the inner accretion zone and the outer zone where the star 
cluster wind is formed. In the second case (above the critical luminosity),
catastrophic cooling sets in inside the cluster. In this case the injected
plasma becomes thermally unstable that inhibits a complete stationary solution.

We compared the calculated accretion rates and the BH luminosities with those 
predicted by the classic Bondi accretion theory and found that Bondi's
theory is in good agreement with our results in the case of low mass clusters.
However, it substantially underestimates the accretion rates and BH 
luminosities if the star cluster mechanical luminosity, $L_{SC} \ge 
0.1 L_{crit}$.
\end{abstract}


\section{Introduction}

Intensive studies of active galactic nuclei (AGNs) in the optical, infrared 
(IR) and X-ray regimes during the last decade, have  unveiled  
the presence of massive starbursts around the central supermassive black 
hole (BH) in a number of Seyfert galaxies. 
For example, Rodr\'\i{}guez Espinosa et al. (1987) found that the 25, 60 and 
100~$\mu$m fluxes of classical optically selected Seyfert galaxies are not 
correlated with their ultraviolet (UV) and optical continuum and that 
far-IR colors of the selected galaxies are indistinguishable from 
those  of starburst galaxies. They suggested then that high far-IR luminosities
associated with many Seyfert galaxies indicate  an intrinsic link between
the circumnuclear star formation and the AGN activity. 
Baum et al. (1993) revealed a kiloparsec-scale, diffuse radio emission
in 12 of the 13 observed Seyfert galaxies. They found that the intensity
of the diffuse radio emission correlates with the far-IR luminosity of the 
host galaxy, suggesting that this emission is generated in the galactic
superwind plasma driven by the nuclear starburst along the minor axis of 
the galaxy, and  
claimed that circumnuclear starbursts and starburst driven outflows may be 
intrinsic to many Seyferts although their relative strengths may vary from 
galaxy to galaxy.
 
Levenson et al. (2001) presented X-ray imaging and spectroscopy of a sample 
of 12 Seyfert 2 galaxies and concluded that in order to fit the observed
X-ray spectra it is required to combine the power-law Seyfert component with 
a thermal starburst emission. Jim\'enez-Bail\'on et al. (2005) presented  
XMM-Newton and Chandra observations of the Seyfert 2 galaxy NGC 1808. They 
found the hard X-ray emission associated with an unresolved nuclear 
sources whereas the soft emission  is dominated by a thermal component 
associated with an extended starburst. 

Terlevich et al. (1990) suggested to use the stellar CaII triplet absorption
feature in the IR continuum as a direct indicator on the presence of young
unresolved stellar population in the nuclear regions of Seyfert galaxies. 
Heckman et al. (1997) and Gonz\'alez Delgado et al. (1998) found absorption 
line features associated with photospheres of O and B stars and their stellar
winds in the ultraviolet and optical spectra of four Seyfert 2 galaxies: 
Mrk 477, NGC 7130, NGC 5135 and IC 3639 and thus presented direct 
evidence for the existence of nuclear starbursts in these galaxies.  
They found that the size of the nuclear starbursts in these galaxies ranges 
from several tens to a few hundred parsecs. The co-existence of W-R features 
in the optical and Ca II triplet in the near-IR part of the spectra, implies 
either a continuous star formation during more than $\sim 10$~Myr or two 
stellar generations with ages about  5-6~Myr and 10-20~Myr, respectively. 
Such starbursts are likely to drive the high-velocity outflows detected in 
the above and in Seyfert 2/starburst ultra luminous infrared 
galaxies (Gonz\'alez Delgado et al. 1998; Rupke et al. 2005).

On the other hand, compact, bright stellar clusters or nuclear star clusters, 
were found in the centers of $\sim 75$\% of local spirals and Virgo dwarf 
elliptical galaxies (B\"oker et al. 2002; C\^ot\'e et al. 2006). Their radii 
(a few parsecs) are similar to those of globular clusters, however they are 
1 - 2 orders of magnitude brighter, more massive and may have complicated 
star formation histories with several episodes of star formation (Walcher et 
al. 2006).
Ferrarese et al. (2006) claimed that massive galaxies (M$_{gal} > 
10^{10}$\Msol) host supermassive BHs whereas less massive  galaxies host only 
nuclear clusters. However Seth et al. (2007) presented evidences on
the presence of super massive black holes in $\sim 25$\% of galaxies which 
host nuclear star clusters. More than half of these galaxies ($\sim 15$\%) 
present a mixed AGN-starburst optical spectra and are classified as composite. 

Shlosman \& Begelman (1989), Goodman (2003), Collin \& Zahn (1999), Tan \& 
Blackman (2005) have shown that accretion disks are gravitationally unstable 
outside of $r \sim 10^{-2}$~pc and must fragment into self-gravitating clumps 
that eventually form stars. It was suggested that star formation reduces the 
rate of accretion and thus the luminosity of the central supermassive black 
hole (BH) by removing mass from the accretion flows and also due to 
radiative heating of the accretion discs. However non of these models 
took into consideration the negative feedback provided by the mechanical 
energy of the central starburst on the accretion flow.
  
Thus circumnuclear star formation occurs at different space scales around 
the SMBH in many AGN galaxies. Here we note that the mechanical power of 
young nuclear starbursts might prevent through the cluster winds the 
accretion of interstellar matter from the bulges and disks of their host 
galaxies onto the central BHs. In such cases  the BHs are fed with the matter 
injected by numerous stellar winds and SNe explosions that result from the 
multiple evolving sources. This implies that nuclear starbursts 
must strongly affect and perhaps even control the power of the central BH. 
In fact, it  may be the dominant factor to be consider in order to understand
the physics and  relative contributions of the BH and starburst
activity to the energy budget of the composite AGN/starburst galaxies. 
L\'ipari \& Terlevich (2006) have incorporated different ingredients of
this physics into their evolutionary unification scenario which seems to 
be able to explain many properties of AGNs and QSOs. 

The classic spherically-symmetric accretion model (Bondi, 1952; Frank et al. 
2002) should then be modified if one is to apply it to the case of a massive 
BH at the center of a  young stellar cluster. First, it should  take 
into consideration the energy and mass supplied by massive stars within the 
star cluster volume. Second, it should account for radiative losses of energy 
from the hot thermal plasma. Third, the models should incorporate  initial 
and boundary conditions as described by the star cluster wind theory 
(Chevalier \& Clegg, 1985; Canto et al. 2000; Silich et al. 2004; 
Tenorio-Tagle et al. 2007). 

Here we present a self-consistent semi-analytic theory of stationary
spherically-symmetric flows driven by the thermalization of the kinetic
energy supplied by massive stars inside massive stellar clusters which
includes the outflow of the injected matter and its accretion onto a 
central massive BH. The solutions account for proper initial and boundary 
conditions for a variety of stellar clusters and black holes and the impact 
of strong radiative cooling on the dynamics of the thermalized injected matter.

The paper is organized as follows. In section 2 we formulate our model and 
discuss the input physics and major simplifications. In section 3 we present
a set of major equations. Boundary conditions and the selection of the proper
solution from a family of integral curves are discussed in section 4. 
In section 5 we discuss the impact that a central BH provides on the 
star cluster driven flows. We discuss two hydrodynamic regimes separated in 
the star cluster mass vs star cluster radius diagram by a threshold 
line: the BH dominated regime below the threshold line and the radiative 
cooling dominated regime above the threshold line. In section 6 we use our 
model to calculate the accretion rates and BH accretion luminosities and 
compare them to those, predicted in the classical Bondi accretion theory.  
Section 7 summarizes our results and gives our conclusions.

\section{The model}

Following Chevalier \& Clegg (1985) we assume that massive stars are 
homogeneously distributed inside a spherical volume of radius $R_{SC}$ and 
that the mechanical energy deposited through stellar winds and supernovae, 
$L_{SC}$, is thermalized via random collisions of the gaseous streams  from 
neighboring  sources. This results into a high temperature and a high thermal 
pressure that leads to a fast outflow of the injected matter,
while composing a stationary star cluster wind.

In presence of a massive, central black hole, a fraction of the 
deposited matter is to remain bound inside the cluster to eventually fall 
onto the center. This implies that in presence of a BH the stagnation 
point, the point where the expansion velocity, $u_w = 0$ km s$^{-1}$, 
is not at the star cluster center, as it happens if the cluster
does not contain a BH and evolves in the quasi-adiabatic regime 
(see below), but instead is at a distance, $R_{st}$, from the star cluster 
center. The position of the stagnation
point thus becomes an important issue that defines both the upper limit 
for the accretion rate onto the central BH, and the amount of matter that 
the star cluster returns, in the form of a wind, to the interstellar medium.

As it was mentioned by Nulsen \& Fabian (2000), the dissipation of angular
momentum is less of a problem for the hot plasma. Thus we restrict ourselves
to spherically-symmetric solutions, despite realizing that some fraction
of the injected material that remains bound inside the star cluster and
that falls onto the center, eventually forms an accretion disk associated with 
the central black hole. We assume that this occurs at a sufficiently
small radii and neglect in this paper all effects associated with the 
redistribution of the residual angular momentum. 

We do not consider the relativistic effects and stop our 
integration at $r = 3 R_{Sh}$, the last stable orbit around the central 
black hole, where $R_{Sh}=2 G M_{BH}/ c^{2}$ is the Schwarzschild radius, 
$M_{BH}$ is the black hole mass, $c$ is the speed of light and $G$ is the 
gravitational constant. 

Throughout the calculations the parameters of the cluster, and the mass of the 
central black hole, are not allowed to change and thus only stationary 
solutions are discussed.

Figure 1 presents a schematic representation of the internal structure 
of the  bimodal flow induced by the presence of a central BH. In this case 
a fraction of the matter supplied by SNe and stellar winds is captured by 
the central BH and the rest is ejected as a wind into the ambient ISM.
\begin{figure}[!htbp]
\plotone{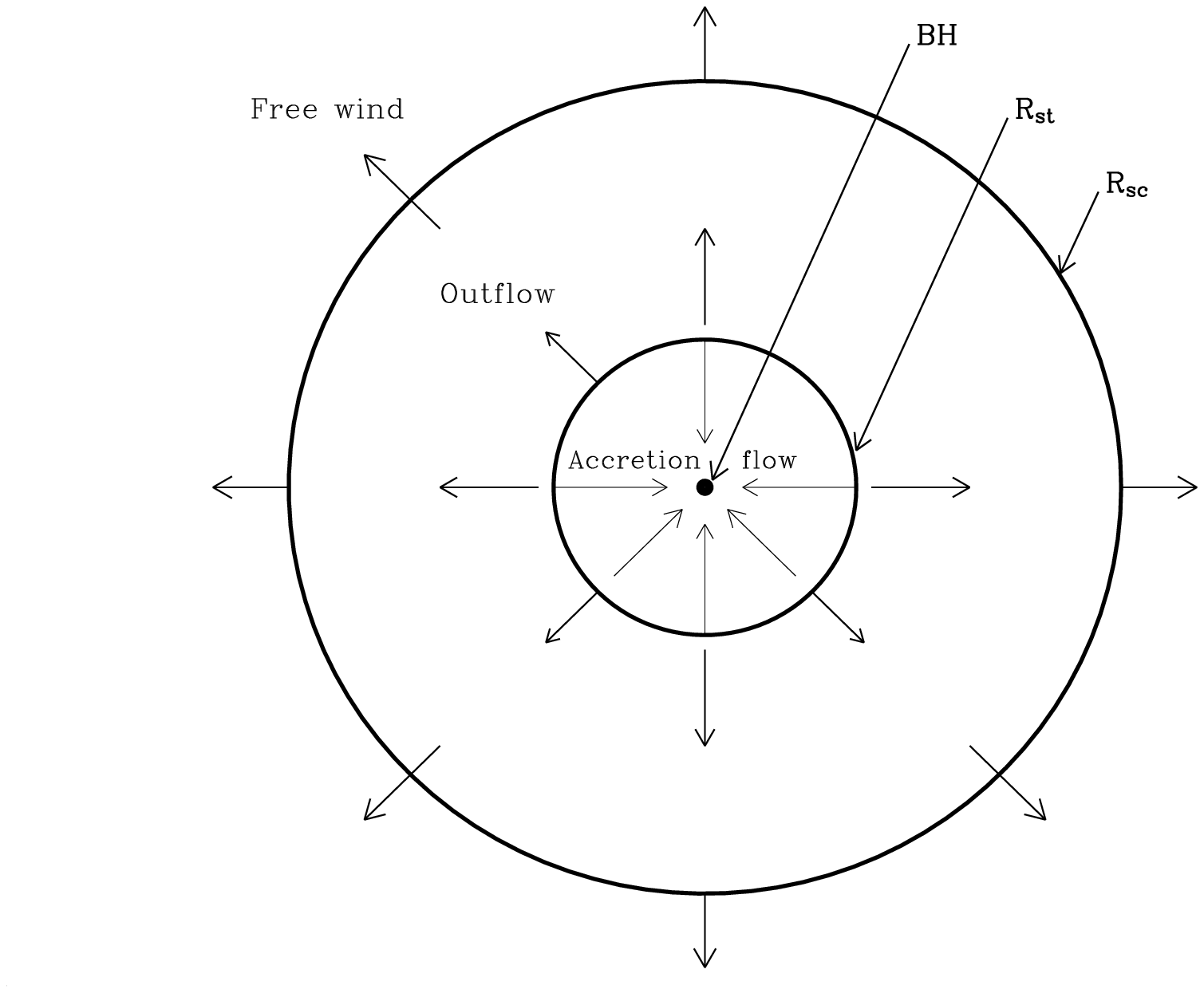}
\caption{The structure of the flow that results from the thermalization 
of the supernova ejecta and stellar winds inside a young stellar cluster,
when a massive black hole is located at the center. The radii of the internal 
and the external circles 
represent the stagnation radius $R_{st}$ and the star cluster radius $R_{sc}$, 
respectively. The arrows indicate the direction of the flow. The black dot at
the center marks the location of the black hole.}
\label{diagram}
\end{figure}

\section{Main Equations}

The hydrodynamic equations for the steady-state, spherically symmetric
flow that results from the energy and mass deposition within young and massive
star clusters with a radius  $R_{SC}$, mass $M_{SC}$ and mechanical
luminosity $L_{SC}$, are (see, for example, Johnson \& Axford, 1971; 
Chevalier \& Clegg, 1985; Cant{\'o} et al. 2000; Silich et al. 2004):
\begin{eqnarray}
      \label{eq1a}
      & & \hspace{-1.0cm}
\frac{1}{r^2} \der{}{r}\left(\rho u r^2\right) = q_m ,
      \\[0.2cm]
      \label{eq1b}
      & & \hspace{-1.0cm}
\rho u \der{u}{r} = - \der{P}{r} - q_m u - 
           \frac{G \rho [M(r) + M_{BH}]}{r^2} ,
      \\[0.2cm]
     \label{eq1c}
      & & \hspace{-1.0cm}
\frac{1}{r^2} \der{}{r}{\left[\rho u r^2 \left(\frac{u^2}{2} +
\frac{\gamma}{\gamma - 1} \frac{P}{\rho}\right)\right]} = q_e - Q -
      \frac{G \rho u [M(r) + M_{BH}]}{r^2},
\end{eqnarray}
and 
\begin{eqnarray}
      \label{eq1d}
      & & \hspace{-1.0cm}
\frac{1}{r^2} \der{}{r}\left(\rho u r^2\right) = 0 ,
      \\[0.2cm]
      \label{eq1e}
      & & \hspace{-1.0cm}
\rho u \der{u}{r} = - \der{P}{r} - 
           \frac{G \rho [M_{SC} + M_{BH}]}{r^2} ,
      \\[0.2cm]
     \label{eq1f}
      & & \hspace{-1.0cm}
\frac{1}{r^2} \der{}{r}{\left[\rho u r^2 \left(\frac{u^2}{2} +
\frac{\gamma}{\gamma - 1} \frac{P}{\rho}\right)\right]} = - Q -
      \frac{G \rho u [M_{SC} + M_{BH}]}{r^2},
\end{eqnarray}
for $r < R_{SC}$ and  $r > R_{SC}$, respectively. $P$, $u$, and 
$\rho$ in equations (\ref{eq1a} - \ref{eq1f}) are the thermal
pressure, the velocity and the density of the thermalized matter. 
The mass and the energy deposition rates per unit volume,
$q_m = 3 {\dot M}_{SC} / 4 \pi R^3_{SC}$ and $q_e = 3 L_{SC} / 
4 \pi R^3_{SC}$, are assumed to be spatially constant inside the star cluster 
and equal to zero if $r > R_{SC}$.
$Q = n_e n_i \Lambda(T,Z)$ is the cooling rate. 
$\Lambda(T,Z)$ is the cooling function, $T$ is the temperature and $Z$ 
is the metallicity of the plasma. $L_{SC} = {\dot M}_{SC} V^2_{A\infty} / 2$,
where $V_{A\infty}$ is the adiabatic outflow terminal speed. $M(r)$ tracks
the distribution of the stellar mass within  the cluster, and $M_{BH}$ is
the mass of the central black hole. We have taken into consideration the 
gravitational pull provided by the central black hole and the star cluster 
and neglected the self-gravity of the reinserted gaseous component.

One can easily integrate the mass conservation equation, both, inside
and outside of the cluster, and rewrite equations (\ref{eq1a} - \ref{eq1f})
in the form:
\begin{eqnarray}
      \label{eq2a}
      & & \hspace{-1.1cm} 
\rho = \frac{q_m r}{3 u}\left(1 - \frac{R^3_{st}}{r^3}\right) \, ,
      \\[0.2cm]     \label{eq2b}
      & & \hspace{-1.1cm}
\der{P}{r} = - \rho u \der{u}{r} - q_m u - 
                 \frac{G \rho (M(r) + M_{BH})}{r^2} \, ,
      \\[0.2cm]     \label{eq2c}
      & & \hspace{-1.1cm}
\der{u}{r}  = \frac{1}{\rho} \frac{(\gamma-1)(q_e - Q) + 
              q_m \left[\frac{\gamma+1}{2}u^2 - \frac{2}{3}
              \left(1 - \frac{R^3_{st}}{r^3}\right) 
              (c_s^2 - V^2_{esc}(r)/4)\right]}{c_s^2 - u^2} \, ,
\end{eqnarray}
within the cluster volume, $r \le R_{SC}$, and 
\begin{eqnarray}
      \label{eq3a}
      & & \hspace{-1.1cm} 
\rho = \frac{{\dot M}_{SC}}{4 \pi u r^2}  \, ,
      \\[0.2cm]     \label{eq3b}
      & & \hspace{-1.1cm}
\der{P}{r} = - \frac{{\dot M}_{SC}}{4 \pi r^2} \der{u}{r} -
                 \frac{G {\dot M}_{SC} (M_{SC} + M_{BH})}{4 \pi r^4 u}
             = - \frac{{\dot M}_{SC}}{4 \pi r^2}
               \left[\der{u}{r} + \frac{V^2_{esc}}{2 r u}\right] \, ,
      \\[0.2cm]     \label{eq3c}
      & & \hspace{-1.1cm}
\der{u}{r}  = \frac{2 u}{r} \frac{2 \pi (\gamma-1) Q r^3 / {\dot M}_{SC} + 
              c^2_s - V^2_{esc}/4}
              {u^2 - c_s^2} \, ,
\end{eqnarray}
in the region $r > R_{SC}$. $c_s = (\gamma P / \rho)^{1/2}$ is the sound 
speed in the hot thermalized ejecta.  
The escape velocity, $V_{esc}$, is:  
$V_{esc} = [2 G (M(r) + M_{BH}) / r]^{1/2}$ if  $r \le R_{SC}$, and 
$V_{esc} = [2 G (M_{SC} + M_{BH}) / r]^{1/2}$ if $r > R_{SC}$, respectively. 

The presence of the BH does not affect the relation between the 
gas number density and the temperature at the stagnation point found in 
Silich et al. 2004:
\begin{equation}
      \label{eq4}
n_{st} = q_{m}^{1/2}\left[\frac{V_{A,\infty}^{2}/2-c_{st}^{2}/(\gamma -1)} 
                   {\Lambda (Z,T_{st})}\right]^{1/2}
\end{equation}
where $V_{A,\infty} = (2q_{e}/q_{m})^{1/2}$ is the adiabatic wind terminal 
speed, $c_{st}$ and $\Lambda(Z,T_{st})$ are the sound speed 
and the cooling function calculated at $r = R_{st}$. One can prove this
result by comparing the derivative of the expansion velocity at the stagnation
point obtained from equation (\ref{eq2c}) with that obtained from equation 
(\ref{eq2a}), and requiring a finite derivative of density at the
stagnation point. Note that Sarazin \& White (1987) obtained similar relation 
from the energy conservation equation in their cooling flow model. 

In all calculations we use the equilibrium cooling function for optically 
thin plasma tabulated by Plewa (1995) and assume that the metallicity of 
the plasma is solar.

\section{Boundary conditions and the appropriate integral curve}

The thermalization of the mechanical energy supplied by massive stars 
within a young stellar cluster, causes a large thermal overpressure
that drives away the injected matter in the form of a
high velocity outflow - the star cluster wind. The smooth transition from 
a subsonic expansion of the high temperature
thermalized ejecta, inside the star cluster volume, to the supersonic free 
wind outflow at $r > R_{SC}$, requires (see equations \ref{eq2c} and 
\ref{eq3c}) the sonic point (the point where the outflow velocity is equal 
to the local speed of sound) to be located at the star cluster surface 
(see Cant{\'o} et al. 2000; Silich et al. 2004). 
Hereafter we will refer to this sonic point as the {\it outer} sonic point.

In the case of stellar clusters with a central black hole, the 
gravitational pull of the BH prevents the escape of the injected matter from 
the central zones of the cluster and thus shifts the stagnation point from 
the star cluster center to a larger radius. In this case all mass continuously
deposited by the cluster inside the central zone, limited by the stagnation 
radius, cannot escape from the gravitational well of the central BH and 
composes the accretion flow. The presence of the central BH results also in 
the existence of the second sonic point, between the stagnation radius and the
star cluster center. To distinguish this sonic point from that at the
star cluster surface we will refer to it as the {\it inner} sonic point.

Thus the stagnation radius defines the upper limit to the accretion rate 
onto the central BH and also the fraction of mass that the cluster returns 
to the ambient ISM. This implies that the major problem that one has to solve 
in order to build a self-consistent hydrodynamic solution for the flow that 
results from the energy and mass deposition by a young stellar cluster with 
a central massive BH is reduced to the calculation of the stagnation radius.

We show below that the proper position of the stagnation
point is defined by the second boundary condition which is similar to that 
in the case of the Bondi accretion with $\gamma=5/3$. Specifically, we 
show that the {\it inner} sonic point must be located at the star cluster 
center. To avoid numerical problems associated with the central singularity 
we will assume that the inner sonic radius coincides instead with the last 
stable orbit associated with the central black hole:
\begin{equation}
\label{eq5}
R_{sonic,in} = 3 R_{Sh} ,
\label{cond}
\end{equation} 
where $R_{Sh}$ is the Schwarzschild radius of the central black hole. 

In order to select a proper integral curve we take a trial stagnation
radius and then select $T_{st}$ from our first boundary condition which
requires the outer sonic point to be located at the star cluster
surface (see Silich et al. 2004; Tenorio-Tagle et al. 2007). We calculate
then the number density of the plasma at the stagnation radius from equation
(\ref{eq4}) and use these $R_{st}, T_{st}$ and $n_{st}$ as initial conditions 
to the backward integration from $R_{st}$ towards the star cluster center.

Figure 2 presents the results of the integration for different values of 
the trial stagnation radius. If the selected $R_{st}$ is too large, the 
backward integration leads to a double-valued, unphysical solution, marked
in Figure 2 by the dashed line. In this case the velocity turnoff point 
(marked by a cross symbol) coincides with the inner sonic point far from the 
cluster center. The turnoff point moves towards $r = 0$ when the considered 
$R_{st}$ is 
smaller. This finally leads to an integral curve (marked in Figure 2 by the 
solid line) that approaches the last stable orbit at $R = 3 R_{sh}$ with the 
sound speed. For even smaller $R_{st}$ the 
stagnation density is larger, the accretion velocity does not reach the 
sonic value at $R = 3 R_{sh}$ and may even go to zero as it is shown in 
Figure 2 (dotted line). 
In order to learn how the inner boundary condition affects the solution,
we have provided several runs with different values of $R_{sonic,in}$.
Figure 3 shows that the stagnation radius is a weak function of
$R_{sonic,in}$ and thus the inner boundary condition does not affect the
solution significantly. We select as the proper solution the integral 
curve which has the second sonic point located at $R_{sonic,in}=3 R_{Sh}$.
\begin{figure}[htbp]
\plotone{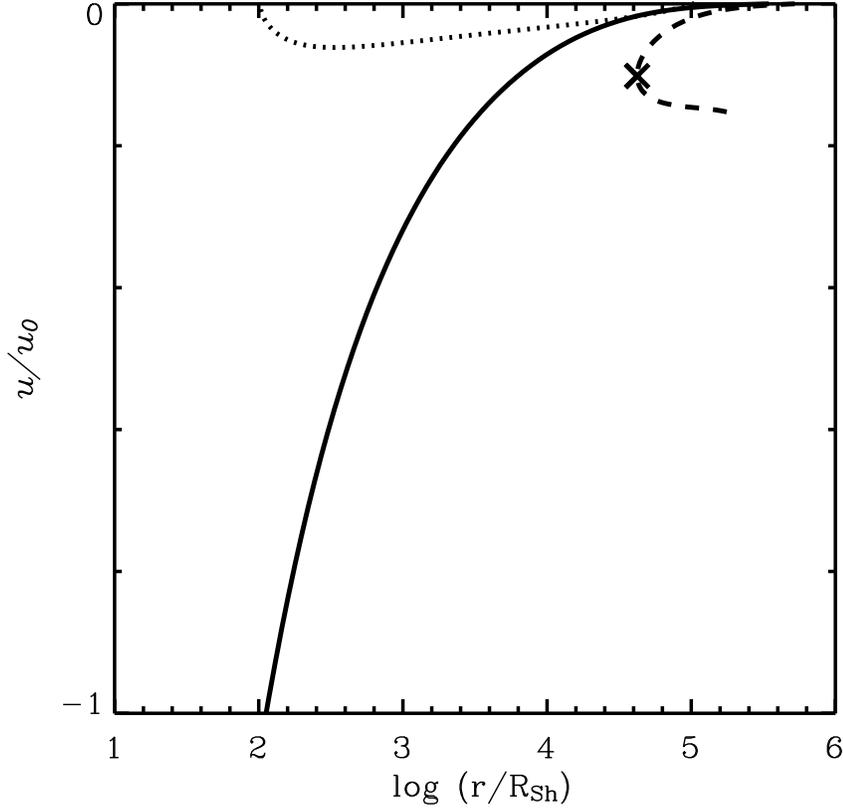}
\caption{Possible integral curves. Three possible integral curves marked by 
dashed, solid and dotted lines correspond to different stagnation radii: 
5 pc, 3.3 pc and 1 pc, respectively. The calculations assumed a star cluster 
mass $M_{SC} = 10^{8}M_{\odot}$, a radius, $R_{SC}=100$ pc, a black hole mass,
$M_{BH} = 10^{8}M_{\odot}$, an adiabatic wind terminal speed, 
$V_{A,\infty}=1500$ km s$^{-1}$, and solar metallicity. The dashed line 
presents the unphysical double-valued solution. The solid line shows the 
selected solution which satisfies for both boundary conditions. The
dotted line presents another unphysical branch of integral curves which
tends towards positive flow velocities around the black hole. 
The normalization velocities are $u_0 = 10^4$~km s$^{-1}$, for dashed and 
solid lines, and $u_0 = 10^2$~km s$^{-1}$ for the dotted line, respectively.}
\label{f2}
\end{figure}
\begin{figure}[htbp]
\plotone{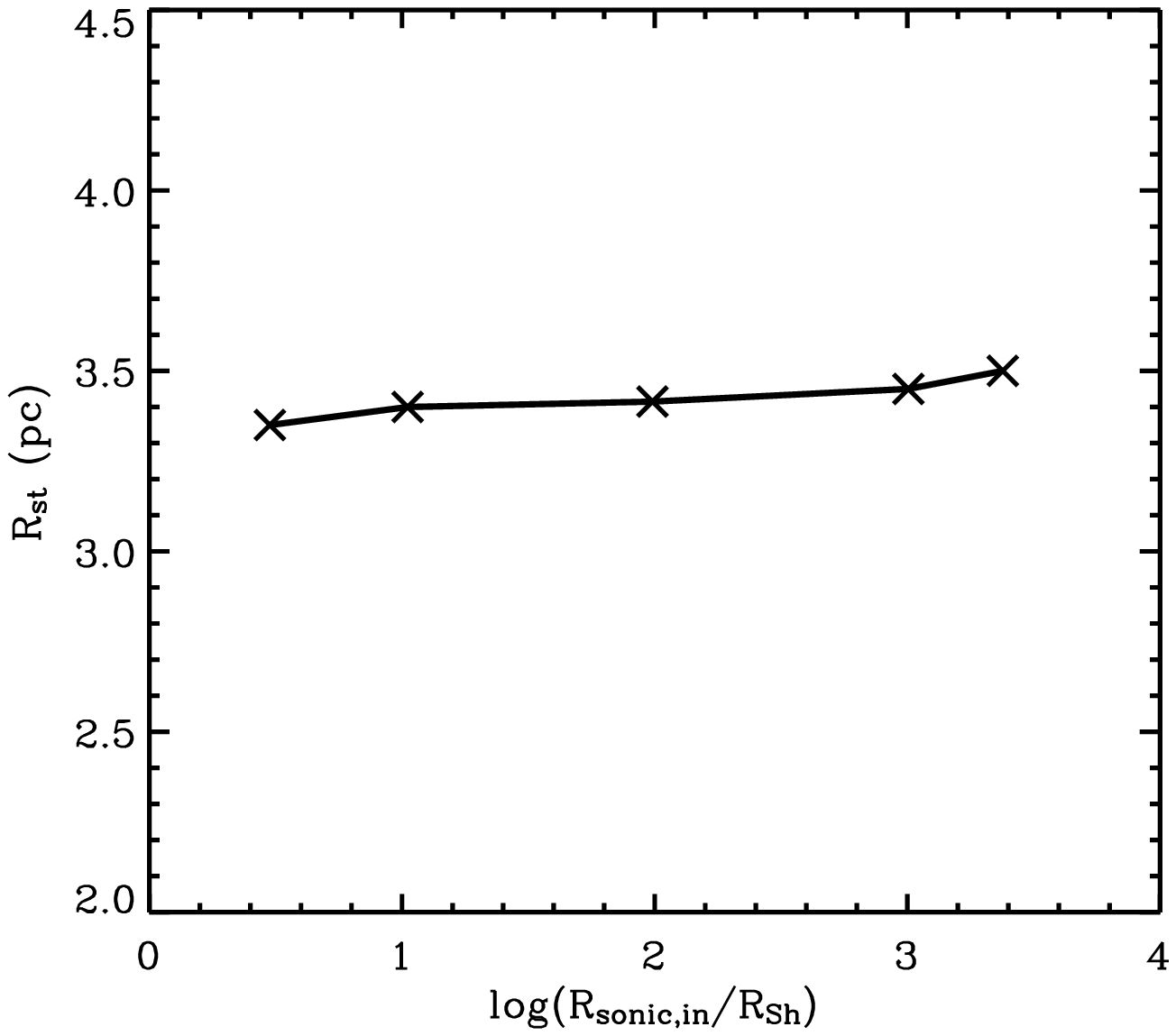}
\caption{The impact of inner boundary condition on the solution. The 
calculations were provided for five different values of inner sonic
radius: $R_{sonic,in} = 3 R_{Sh}; \, 10.5 R_{Sh}; \, 98 R_{Sh}; \,
1009 R_{Sh}$ and $2385 R_{Sh}$. The star cluster and black hole parameters are 
identical to those in Figure 2: $M_{SC} = 10^{8}M_{\odot}$, $R_{SC}=100$ pc, 
$V_{A,\infty}=1500$ km s$^{-1}$ and $M_{BH} = 10^{8}M_{\odot}$.  
Cross symbols represent the results of the calculations.}
\label{f3}
\end{figure}

Figure 4 presents the distribution of the flow variables (velocity, number
density and temperature) for a particular case of a $10^8$\Msol \, black hole 
located 
at the center of a young stellar cluster whose mass $M_{SC} = 10^8$\Msol \, 
and $R_{SC} = 40$pc. In this case the stagnation radius, marked by the inner 
dotted line in panel a, is $R_{st} = 2.7$~pc. At larger radii the velocity 
grows almost linearly to reach the sonic value at the star cluster 
surface. It becomes supersonic outside the cluster and soon reaches the 
terminal value, $V_{\infty}$, somewhat smaller than the adiabatic terminal
speed value, as radiative losses deplete some energy inside the cluster and 
in the free wind region. 

In the region between the stagnation radius and the black hole, the
matter deposited by stellar winds and supernovae composes a stationary 
accretion flow. The absolute value of the velocity grows rapidly in this 
region. However the flow remains subsonic as radiative losses are not able 
to compensate the heating of the in-flowing plasma.
This leads to a rapid increase of temperature (panel c) despite the increase 
in density and thus of cooling of the in-falling matter, as it approaches the 
star cluster center (panel b). 
\begin{figure}[htbp]
\vspace{17.5cm}
\includegraphics{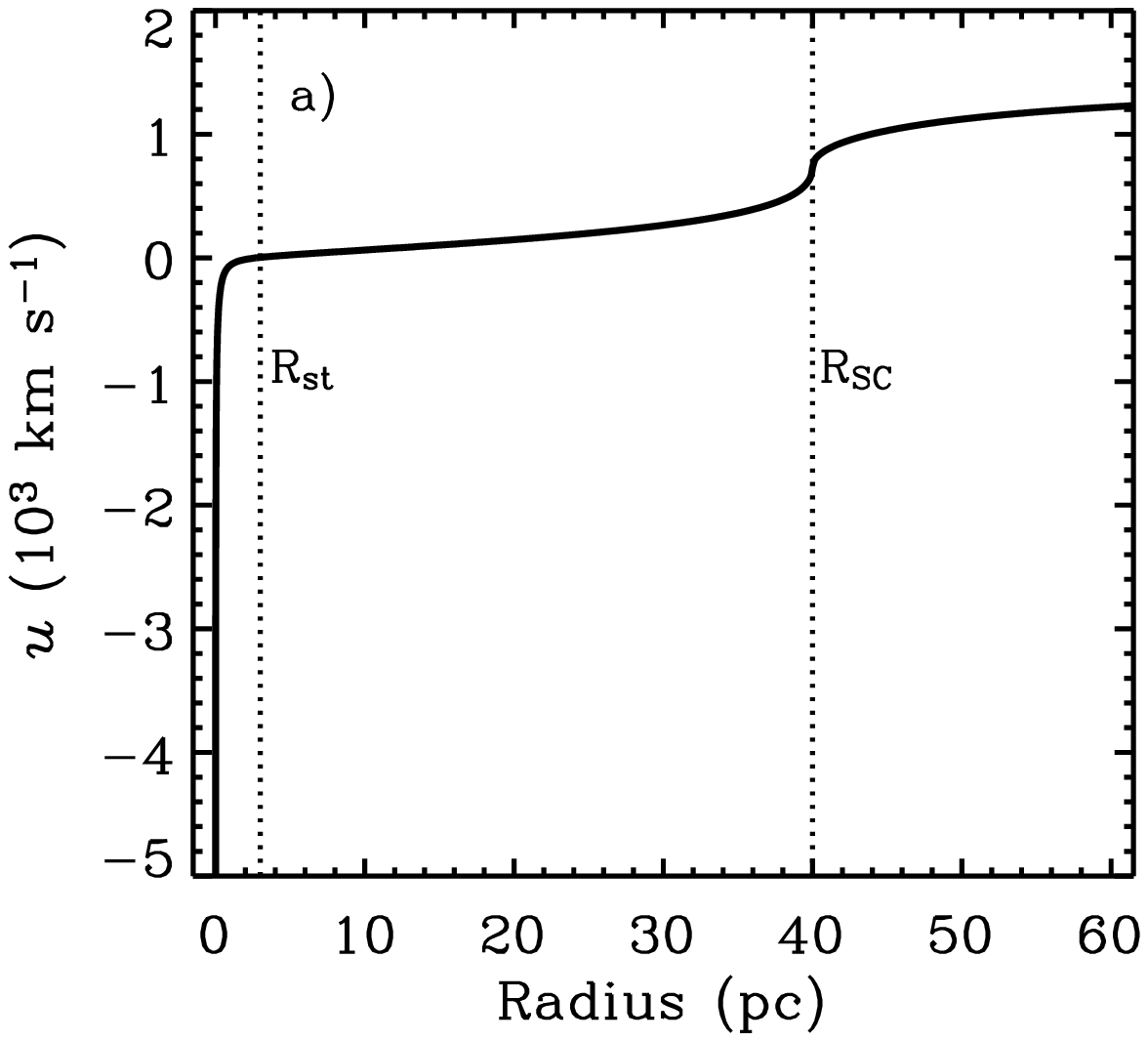}
\includegraphics{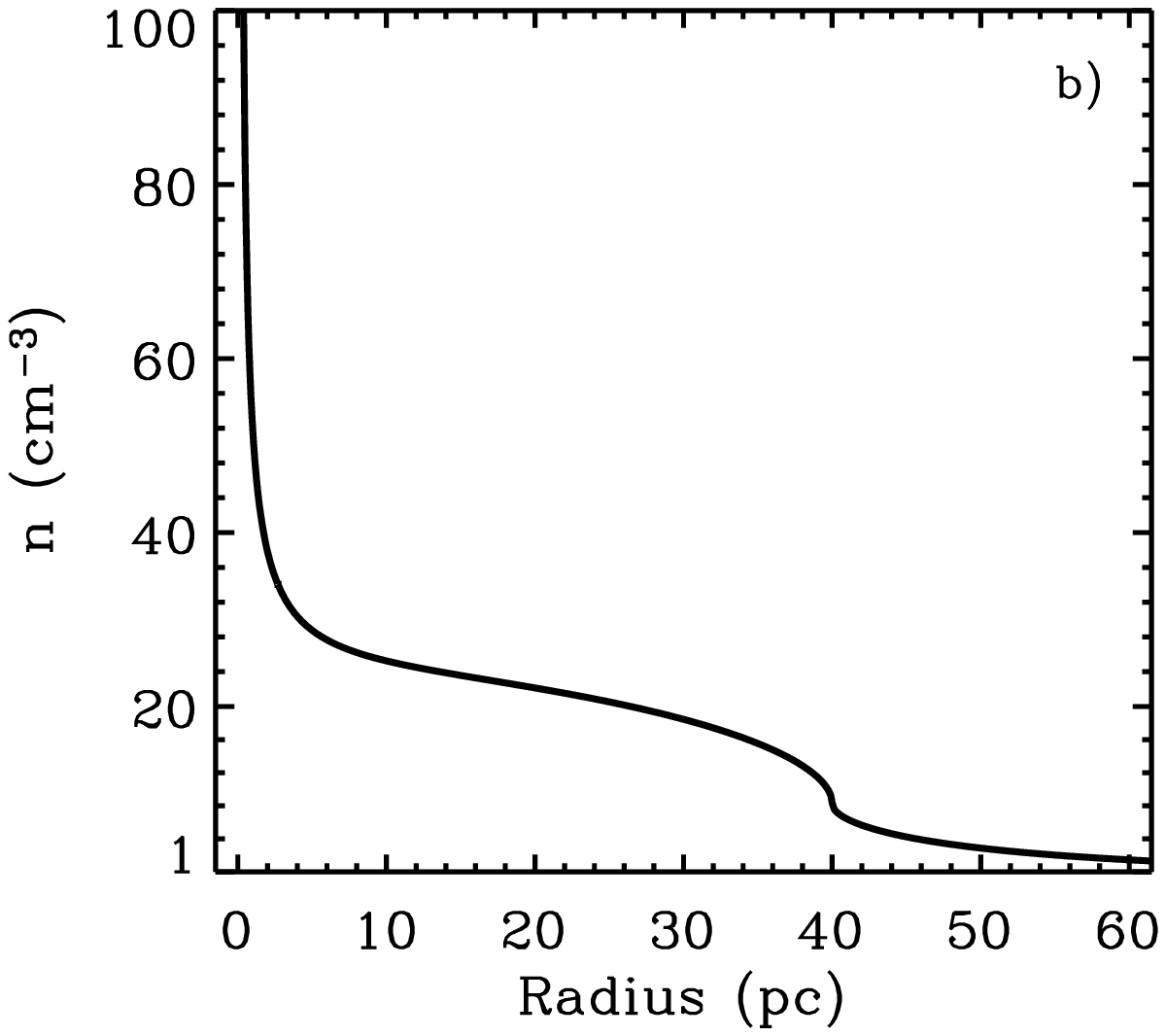}
\includegraphics{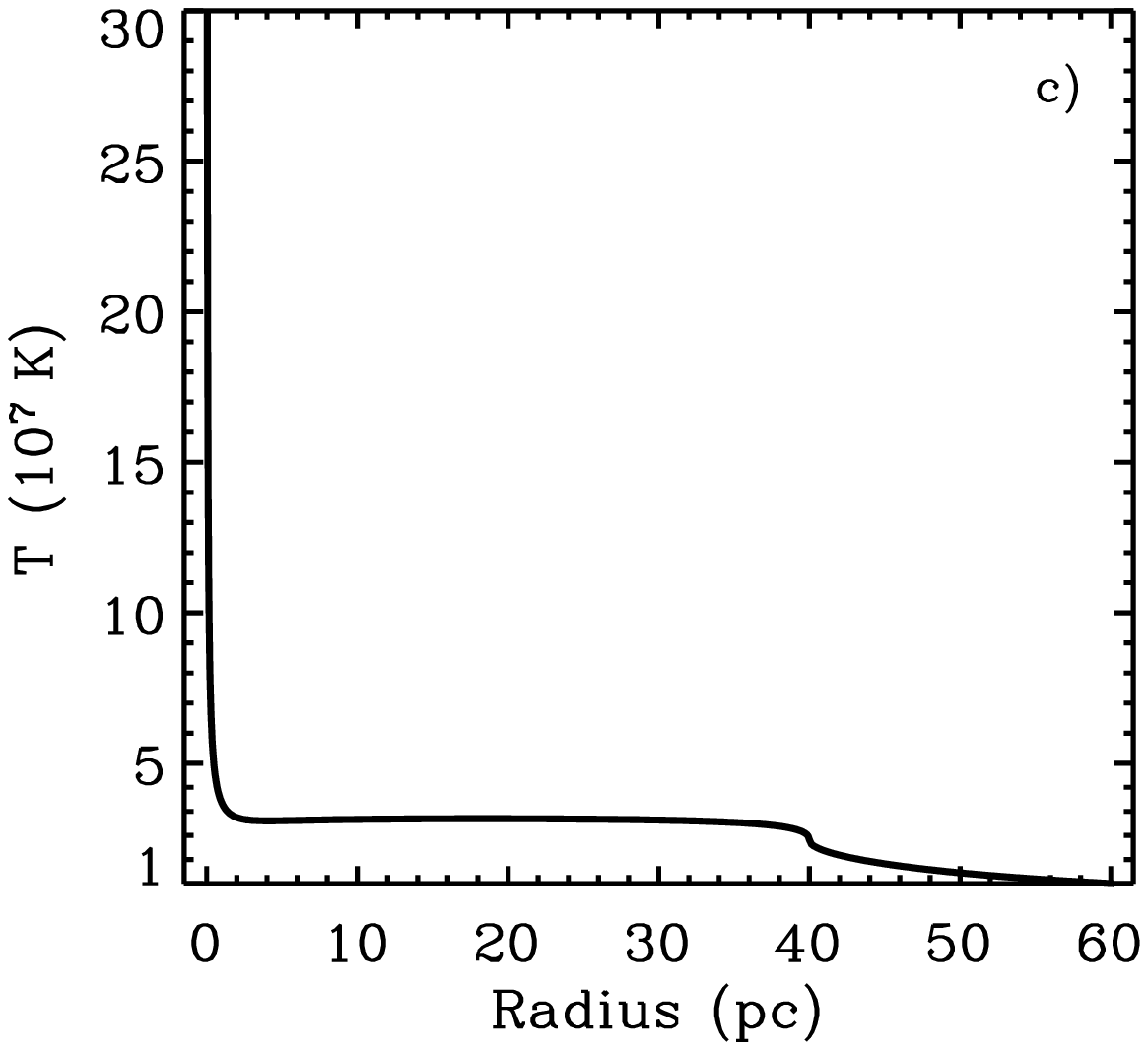}
\caption{The structure of the flow in the case of a young stellar cluster
with a central massive black hole. Panels a, b and c present the distribution
of the flow velocity, number density and temperature, respectively.
The calculations assumed a $10^8$\Msol \, black hole
located at the center of a young stellar cluster whose mass is  
$M_{SC} = 10^8$\Msol \, and $R_{SC} = 40$pc. The adopted mechanical luminosity
of the cluster is $3 \times 10^{42}$erg s$^{-1}$. This value corresponds to 
the average mechanical luminosity of a young stellar cluster with a Salpeter 
initial mass function (Leitherer et al. 1999). It was assumed that 
the adiabatic wind terminal speed is, $V_{A,\infty}=1500$ km s$^{-1}$.
Vertical dotted lines in panel a mark the stagnation radius, $R_{st}$, 
and the star cluster radius, $R_{SC}$, respectively.}
\end{figure}

Figure 5 shows how the location of stagnation point depends on the mass of 
the black 
hole and that of the stellar cluster, when the radius of the cluster 
remains fixed. $R_{st}$ becomes larger when the gravitational pull becomes 
stronger either because one considers more massive black holes or clusters.
\begin{figure}[htbp]
\plottwo{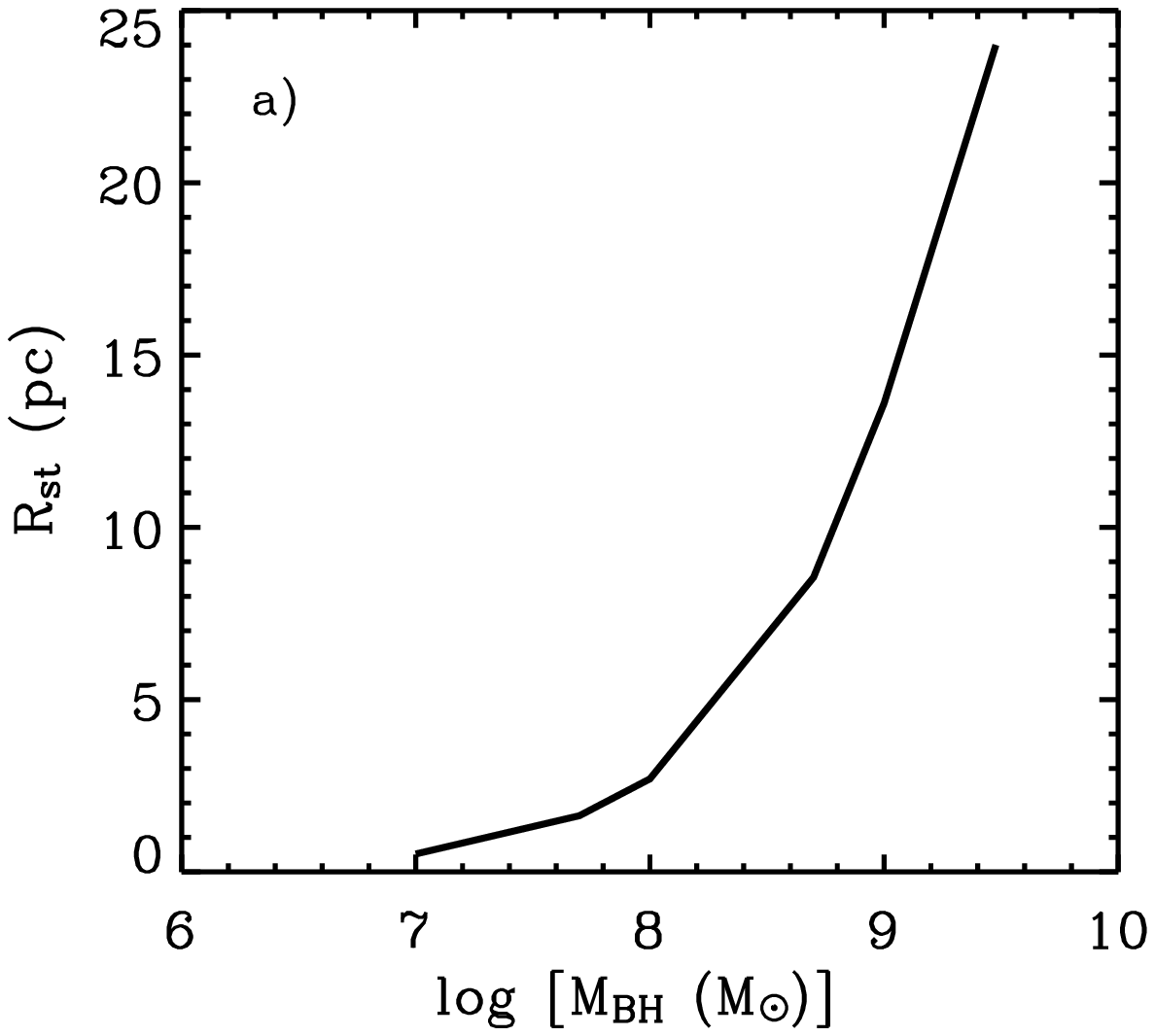}{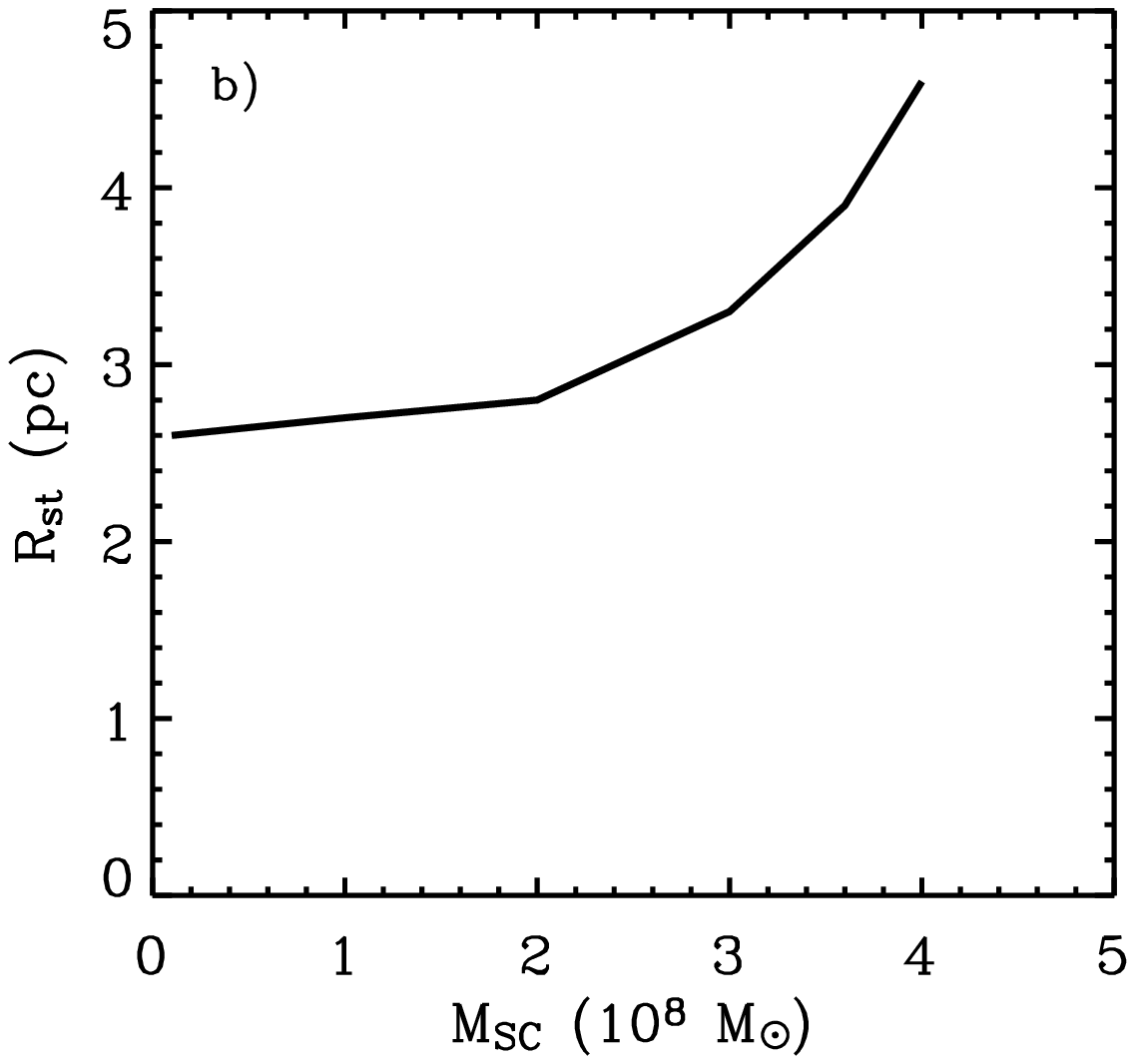}
\caption{The position of the stagnation point. Panel a shows how the 
stagnation radius grows with the mass of the black hole for a fixed 
($10^8$\Msol) star cluster mass, whereas panel b shows the value of the 
stagnation radius as a function of the star cluster mass for a fixed 
($10^8$\Msol) black hole mass. In all cases the radius of the cluster is 
40~pc.}
\end{figure}

\section{The threshold mechanical luminosity}

Silich et al. (2004), Tenorio-Tagle et al. (2007) and W\"unsch et al. (2008)
have thoroughly discussed the impact of radiative cooling on the inner 
structure of star cluster driven flows in the case without BH and 
without accounting for the contribution to
the gravitational field imposed by the cluster. They showed that strong 
radiative cooling changes drastically the pressure gradient in the 
inner zones of compact and massive star clusters and found a critical, 
threshold mechanical luminosity which separates, in the $L_{SC}$ - $R_{SC}$ 
parameter space, clusters with $R_{st} = 0$~pc from those evolving in the 
bimodal, catastrophic cooling regime. In the latter regime, strong radiative 
cooling promotes the  displacement of the stagnation point out of the star 
cluster center and leads to the accumulation of the matter injected inside 
the stagnation volume as it becomes thermally unstable, while the outer zones 
of the cluster drive a stationary outflow. On the other hand, as shown above, 
stellar clusters with a central BH evolve always in a
bimodal regime because the gravitational force goes to infinity when the
distance to the BH goes to zero. This establishes the position of a stagnation
point within the flow and inhibits the escape of the hot plasma
from the cluster inner zones. The implication is thus that the two physical 
processes here considered: the gravitational pull from the central massive BH 
and strong radiative cooling, both promote the existence of a stagnation 
point within the flow and thus both compete in defining the final hydrodynamic 
solution. 

In order to calculate the threshold energy,
we choose the stagnation temperature which leads to the maximum thermal 
pressure at the stagnation point. This is defined by the condition that 
${\rm d} P_{st} / {\rm d} T_{st} = 0$ which, together with equation
(\ref{eq4}), yields
\begin{equation}
      \label{eq.7a}
\left(\frac{V^2_{A\infty}}{2} - \frac{c^2_{st}}{\gamma - 1}\right)
\left(1 - \frac{T_{st}}{2 \Lambda} \der{\Lambda}{T_{st}}\right) -
\frac{1}{2} \frac{c^2_{st}}{\gamma - 1} = 0 .
\end{equation}
Then we iterate the stagnation radius and the star cluster mechanical 
luminosity until both boundary conditions are fulfilled, e.g. until the 
outer sonic point accommodates at the star cluster surface and the inner one
reaches the $3 R_{Sh}$ radius. 

Figure 6 displays the threshold mechanical luminosity calculated for
stellar clusters of different radii which contain black holes of different
masses at their centers. 
The gravitational pull of the BH does not affect too much the value of the
threshold luminosity in the case of large clusters. Although in all of these
cases the critical luminosity is several times smaller than in 
the case without a BH (see Figure 6).

The gravitational pull of the BH becomes progressively more important for 
more compact clusters. The threshold lines turn up when the considered BH 
mass becomes comparable to the critical cluster mass (see Figure 6).
In the case of even more compact clusters, cooling cannot compete with gravity.
The position of the stagnation point is defined then by the mass of the BH and
the stationary solution exists regardless of the cluster mass. Perhaps the
only limitation here arises when $R_{st}$ becomes larger than $R_{SC}$.
This implies that the critical luminosity does not 
exist below $R_{SC,crit}$. The critical radii, $R_{SC,crit}$, depend on 
the mass of the BH and are marked in Figure 6 by thin vertical lines for 
different values of the BH mass.
\begin{figure}[htbp]
\plotone{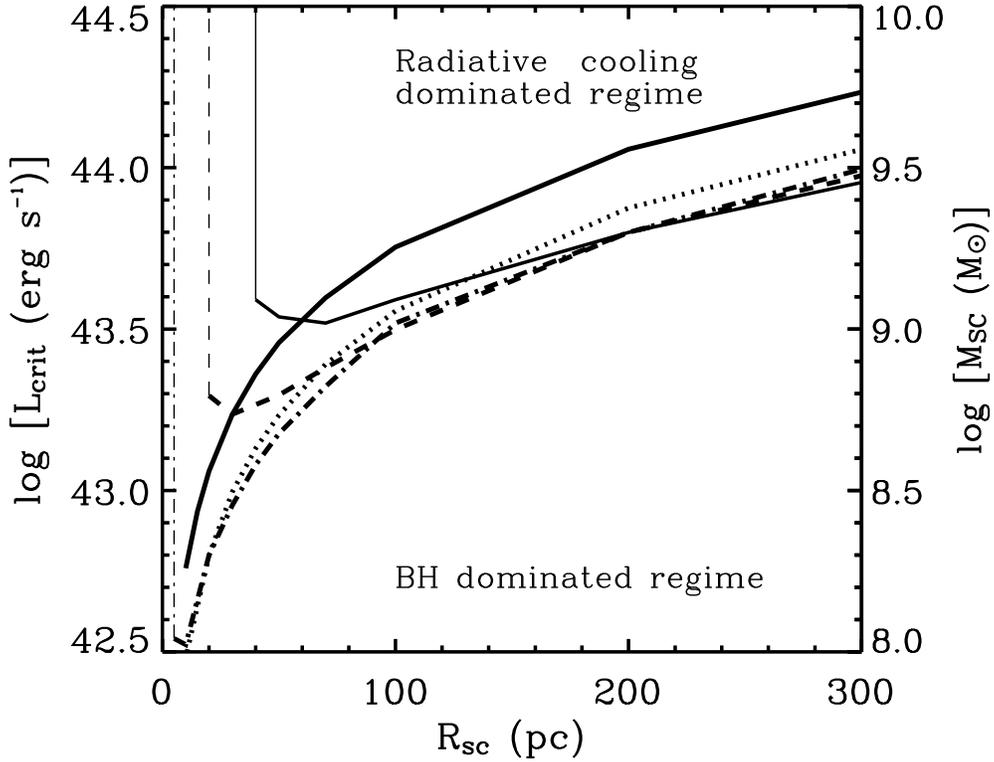}
\caption{The threshold mechanical luminosity. The thick solid line marks 
the threshold luminosity for cases without a central black hole. The 
dotted, dash-dotted, dashed and thin solid lines display the threshold 
luminosity for clusters with $10^7$\Msol \,  $10^8$\Msol \,  $5 \times 
10^8$\Msol \, and $10^9$\Msol \, black holes, respectively. Thin vertical
lines mark critical radii, $R_{SC,crit}$. Cooling cannot compete with 
gravity if the star cluster radius, $R_{SC} < R_{SC,crit}$. In this parameter 
space gravity defines the stagnation radius regardless of the star cluster 
mass and the critical luminosity does not exist. It was assumed an adiabatic 
wind terminal speed equal to 1500~km s$^{-1}$ and solar metallicity.}
\end{figure}

If the cluster does not contain a BH and its gravitational pull is negligible,
the stagnation radius is defined by the excess mechanical
luminosity, $L_{SC}$, over the threshold value (W\"unsch et al. 2007): 
\begin{equation}
      \label{eq6}
\frac{R^3_{st}}{R^3_{SC}} = 1 - \left(\frac{L_{crit}}{L_{SC}}\right)^{1/2} .
\end{equation}

From this it is clear that the impact of radiative
cooling on the inner structure of the flow is to become progressively more
important for star clusters with larger mechanical luminosities (proportional 
to the star cluster mass). For low mass clusters, the gravitational pull of 
the black hole is to dominate and define the position 
of the stagnation point, but radiative cooling is to become an important 
factor when one considers more massive clusters. 

We expect that above the threshold line the internal structure of the flow
would present three distinct zones, as displayed in Figure 7. The outer zone 
would conform a quasi-stationary outflow. This runs from an outer stagnation 
radius $R_{st,cool}$, set by strong radiative cooling, to meet the sound 
speed at the cluster surface. The central region will enclose the accretion 
flow onto the BH. The outer boundary of this flow would be the stagnation 
radius defined by the BH ($R_{st,BH}$). Between these two stationary flows  
the thermalized ejecta is thermally unstable and rapidly decays into two 
phases: a hot plasma whose parameters are similar to those found at the outer 
stagnation point, and a collection dense clouds which result 
from the thermally unstable plasma and are completely or partially 
photo-ionized by the stellar UV and BH hard radiation (W\"unsch et al. 2008). 
These clouds can fall onto the BH or become gravitationally unstable and 
support some level of star formation inside such a cluster. Numerical 
calculations and a thorough discussion of this regime will be the subject 
of a forthcoming communication. 
\begin{figure}[htbp]
\plotone{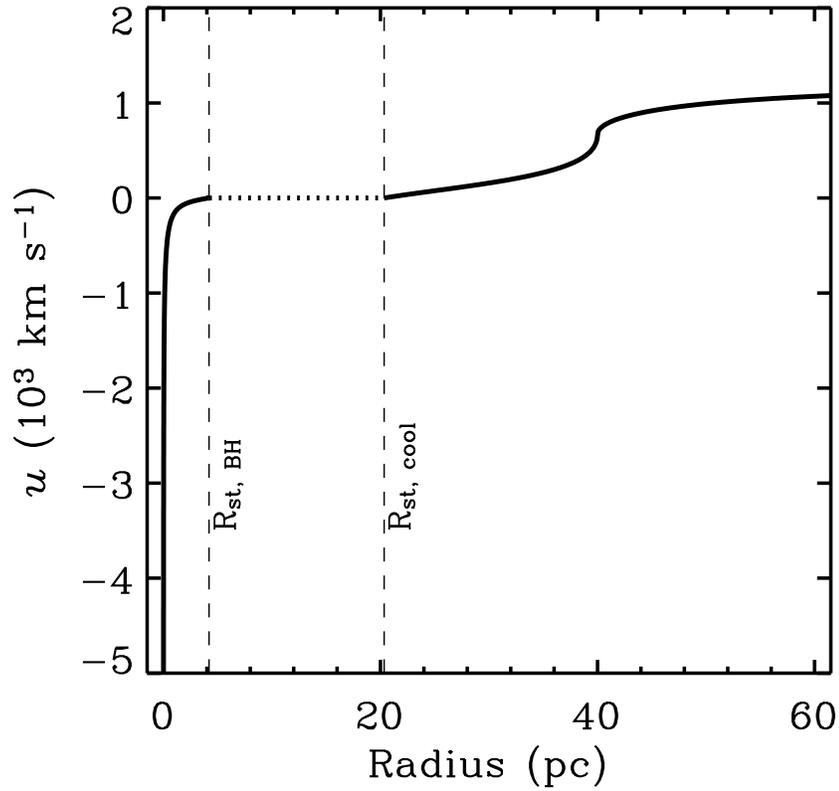}
\caption{The expected velocity pattern of the flow in the case of clusters 
located above the threshold line. The stationary solution does not exist in 
this case. Positions of the inner and outer stagnation points are defined by 
the gravitational field of the central BH and strong radiative cooling, 
respectively. The intermediate zone is thermally unstable.} 
\end{figure}

\section{Accretion rates and BH luminosities}

For clusters below the threshold line, the mass deposited by stellar winds 
and supernovae explosions inside 
$R_{st}$ is not able to escape the stagnation volume and in the stationary
regime has to fall to the center providing fuel to the central BH.
Thus in the stationary regime the stagnation radius, $R_{st}$,
defines the accretion rate onto the central object:
\begin{equation}
\label{eq8}
\dot M_{acc}=\dot M_{sc}\left(\frac{R_{st}}{R_{sc}}\right)^{3} .
\end{equation}

In this respect it is instructive to compare our theory with the classic
spherically-symmetric accretion theory (Bondi, 1952; see Frank et al. 2002 and 
references therein). If the polytropic index $\gamma=5/3$, the Bondi 
accretion rate is (Frank et al. 2002):
\begin{equation}
\label{eq9}
\dot M_{B}=\pi G^{2}M_{BH}^{2}\frac{\rho_{ISM}}{c_{ISM}^{3}},
\end{equation}
where $\rho _{ISM}$ and $c_{ISM}$ are the density of the ISM and the speed 
of sound at infinity, respectively. 
We associate these quantities with Chevalier \& Clegg's central values
(see Cant\'o et al. 2000):
\begin{eqnarray}
      \label{eq10a}
      & & \hspace{-1.1cm} 
\rho_c = \frac{2}{4\pi A}\frac{L_{sc}}{R_{sc}^{2}V_{A,\infty}^{3}},
      \\[0.2cm]     \label{eq10b}
      & & \hspace{-1.1cm}
c_c = \left(\frac{\gamma-1}{2}\right)^{1/2} V_{A,\infty} ,
      \\[0.2cm]     \label{eq10c}
      & & \hspace{-1.1cm}
A = \left( \frac{\gamma -1}{\gamma +1}\right)^{1/2}
  \left( \frac{\gamma +1}{6\gamma +2}\right)^{(3\gamma +1)/(5\gamma +1)} .
\end{eqnarray}

Figure 8 compares our semi-analytic results (equation \ref{eq8}) with Bondi 
accretion rates (equation \ref{eq9}). The calculations were performed for a 
set of clusters which contain a $10^8$\Msol \, BH at the center, all clusters
have the same radius of $R_{SC} = 40$~pc. It was assumed that the adiabatic 
wind terminal speed is $V_{A\infty} = 1500$~km s$^{-1}$ and a plasma with 
solar metallicity. In order to relate the mechanical luminosity of the 
stellar cluster to the corresponding star cluster mass we have used a 
relation which approximates the results of the Starburst 99 synthesis model for
young stellar clusters (Leitherer et al. 1999):
\begin{equation}
\label{eq11}
L_{SC} = 3 \times 10^{40} \left(\frac{M_{SC}}{10^6\Msol}\right) \, erg \, 
s^{-1} .
\end{equation}
The star cluster mechanical luminosity has been normalized to the critical
value derived in the previous section. Thus when $L_{SC} / L_{crit} \ll 1$,
clusters evolves in the quasi-adiabatic regime, whereas when
$L_{SC} / L_{crit} \to 1$ the star cluster parameters approach the threshold 
values. The semi-analytic results are marked by the cross symbols and the 
accretion rates predicted by the Bondi equation are shown by the solid line.

Figure 8 shows that well below the threshold line (when $L_{SC} / L_{crit}
\le 0.1$) Bondi's formula (equation \ref{eq9}) is in good agreement with our 
numeric results.
\begin{figure}[htbp]
\plotone{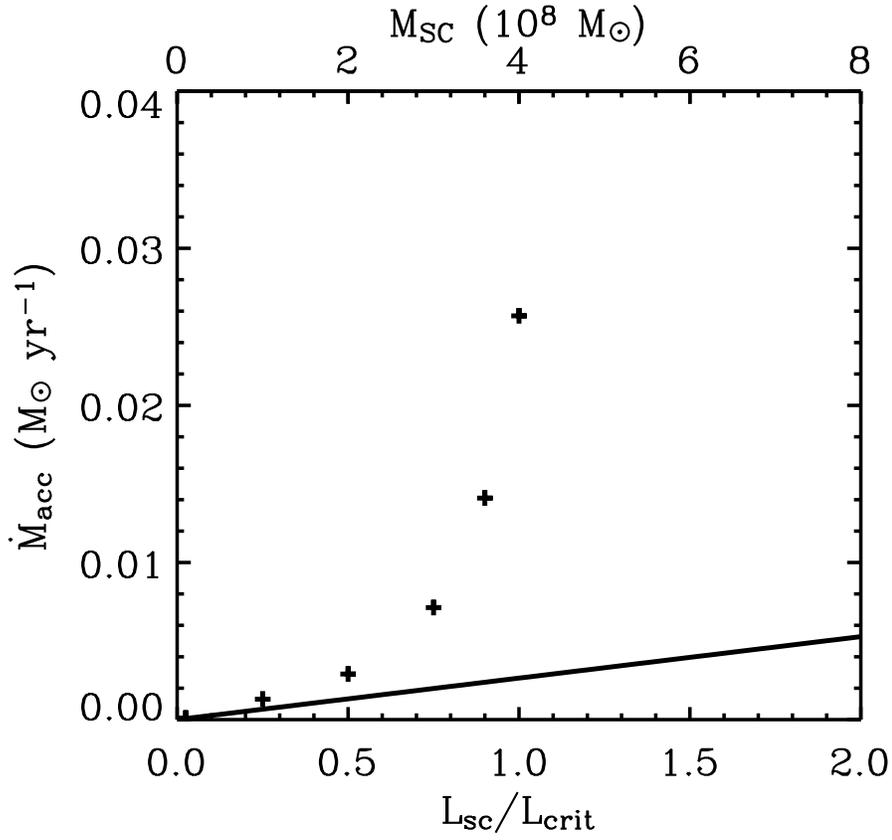}
\caption{The comparison of the approximate analytic formulae with
numeric accretion rates. Cross symbols present the numeric results
for a $10^8$\Msol \, BH embedded into stellar clusters of different masses,
all having radius of $R_{SC} = 40$~pc. Solid line displays the accretion 
rates calculated in the Bondi approximation with the adiabatic wind input 
parameters.}
\end{figure}
For such clusters one can approximate the spherically-symmetric
accretion rate onto a BH located at the center of a young massive cluster
with Bondi's expression:
\begin{equation}
\label{eq13}
\dot M_{acc} = \pi G^{2}M_{BH}^{2}\frac{\rho_{c}}{c_{c}^{3}} ,
\end{equation}
where $\rho_{c}$ and $c_{c}$ are taken from equations (\ref{eq10a}) - 
(\ref{eq10c}). One can then use equation (\ref{eq13}) in order to obtain 
an analytic expression for the stagnation radius in this ($L_{SC} \ll 
L_{crit}$) parameter space. Indeed, the mass accretion rate is
\begin{equation}
\label{eq14}
\dot M_{acc} = 4 \pi q_m R^3_{st} / 3 ,
\end{equation}
where the mass deposition rate per unit volume, $q_m$, is
\begin{equation}
\label{eq15}
q_{m} = \frac{3}{4\pi}\frac{\dot M_{sc}}{R_{sc}^{3}} = 
        \frac{6}{4\pi}\frac{L_{sc}}{V_{A,\infty}^{2}R_{sc}^{3}} .
\end{equation} 
Combining equations (\ref{eq14}) and (\ref{eq15}) with equation (\ref{eq13}),
one can obtain:
\begin{equation}
\label{eq16}
R_{st} = \left(\frac{\pi G^2 V^2_{A,\infty} M^2_{BH} \rho_c}
         {2 L_{SC} c^3_c}\right)^{1/3} R_{SC} .
\end{equation} 

For star clusters whose mechanical luminosities, $L_{SC}$, are comparable to
the threshold value, $L_{crit}$, the calculated accretion rates exceed 
substantially those predicted by Bondi's equation. In this case one has 
to use the semi-analytic model in order to find the size of the stagnation 
zone and then equation (\ref{eq8}) in order to derive the accretion rate. 

Having the accretion rates and adopting the normal accretion efficiency, 
$\eta = 0.1$, one can calculate the BH luminosity. Figure 9 compares
the calculated BH luminosities with the Eddington limit, $L_{Edd} = 1.3 
\times 10^{38} M_{BH}/M_{\odot}$~erg s$^{-1}$, for star clusters evolving
in different hydrodynamic regimes. $L_{BH}$ grows in 
the case of more compact clusters. However, it remains well below the 
Eddington limit even when the star cluster mechanical luminosity reaches
the threshold value (Figure 9, panel a). The BH luminosity can approach 
the Eddington limit (see Figure 9, panel b) in the case of very compact 
clusters whose radii are smaller than the critical values, $R_{SC,crit}$, 
marked in  Figure 6 by vertical lines. This puts a limit for our model in the 
case of very compact clusters.
\begin{figure}[htbp]
\plottwo{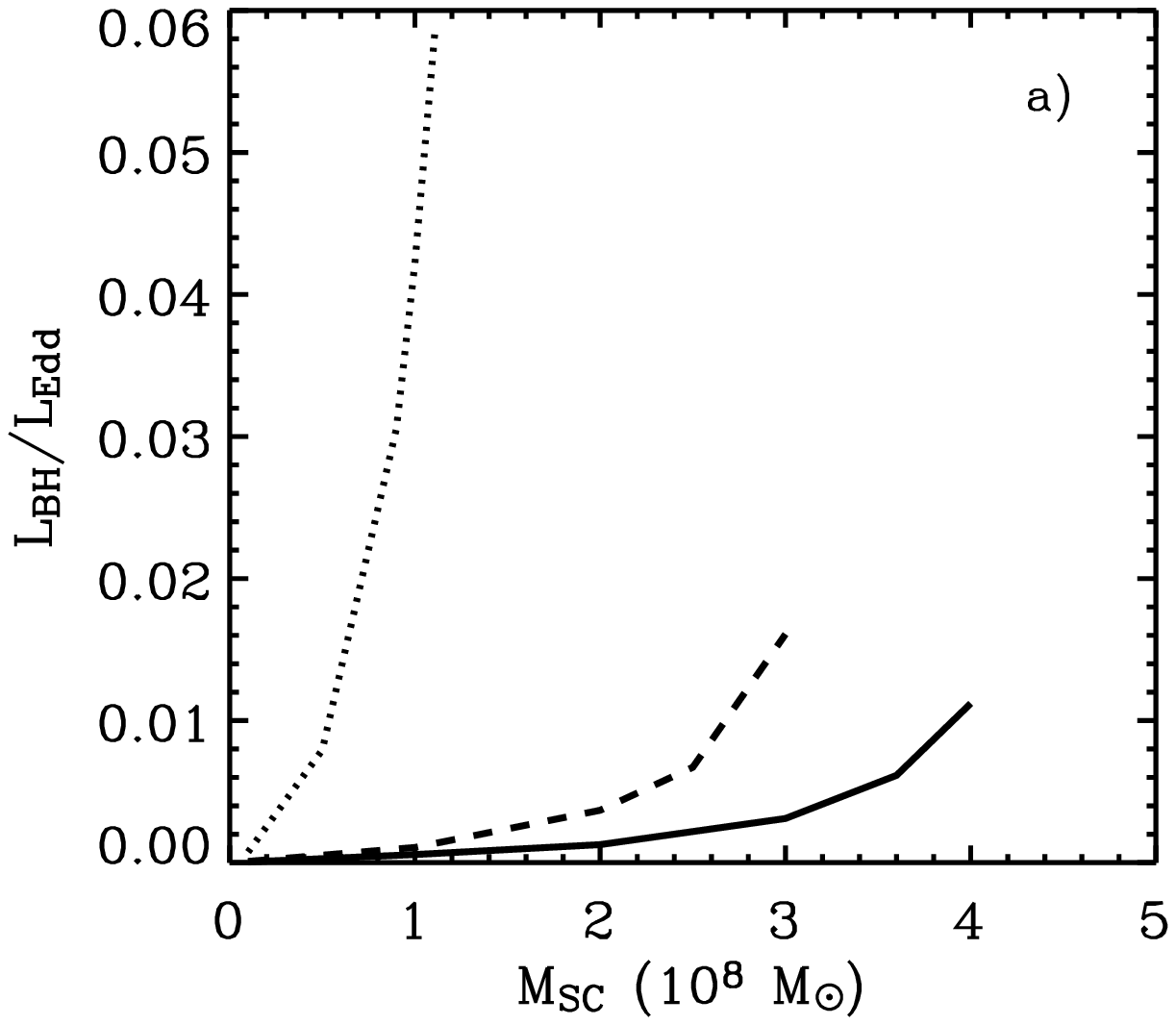}{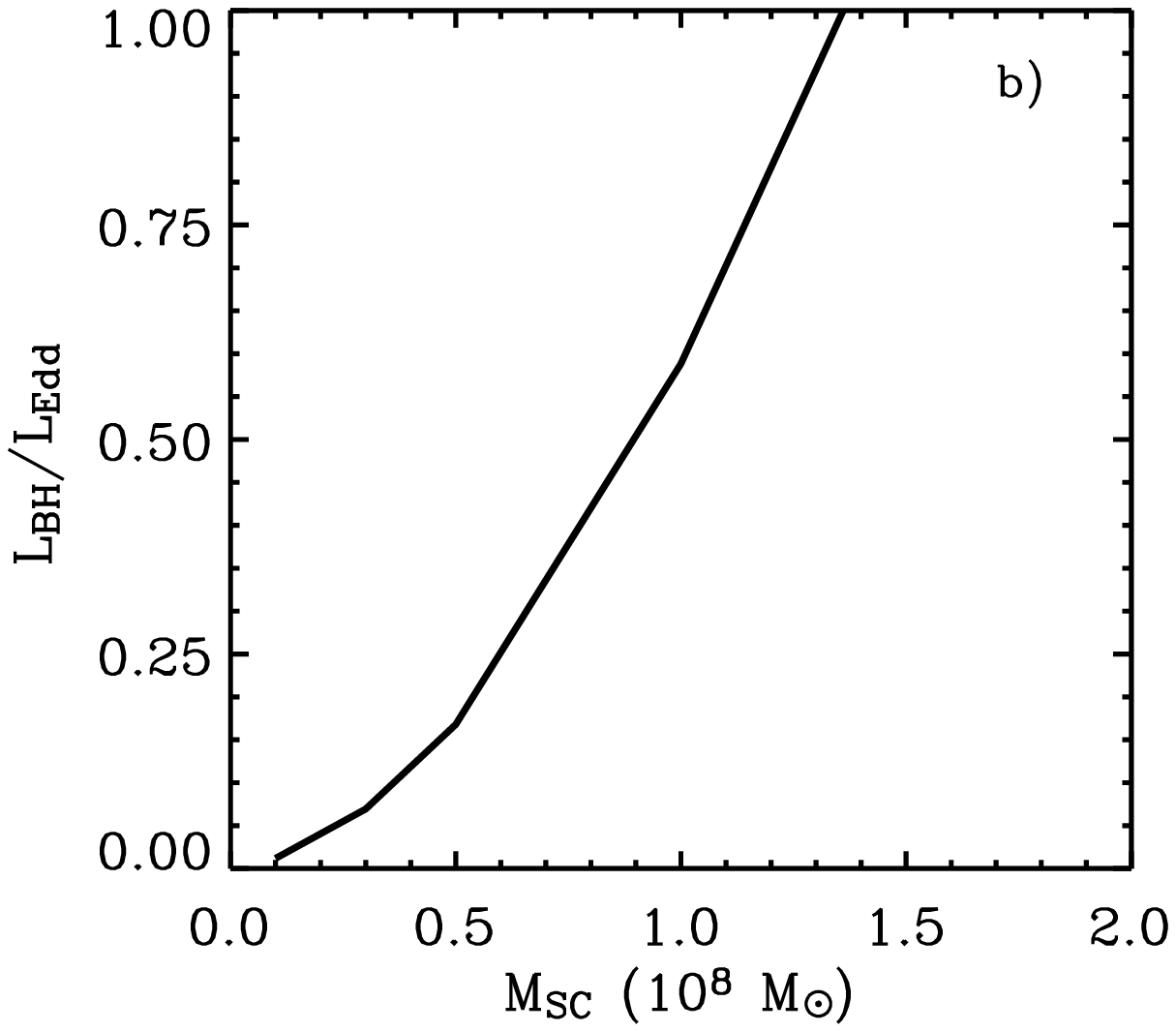}
\caption{The BH accretion luminosity. Panel a presents the accretion 
luminosity of a $10^8$\Msol \, BH embedded into different star clusters.
Solid, dashes and dotted lines display the results of the calculations
for star clusters whose radii are $R_{SC} = 40$~pc, $R_{SC} = 30$~pc
and $R_{SC} = 10$~pc, respectively. The last point on every line
displays the BH accretion luminosity when the star cluster reaches
the threshold line. Panel b presents the accretion luminosity of 
a $10^8$\Msol \, BH embedded into very compact ($R_{SC} = 3$~pc)
star clusters, all located on the vertical dot-dashed line 
in Figure 6.}
\end{figure}

\section{Conclusions}

We have developed a self-consistent, stationary solution for spherically 
symmetric accretion flows which are formed inside young stellar clusters
with a central supermassive BH. 

We have shown that the thermalization of the kinetic energy released by
massive stars inside young stellar cluster results in a bimodal solution
which presents an accretion of the injected matter onto a central BH in
the inner zones of the cluster, and the ejection of the deposited matter
from the outer zones of the cluster in the form of the fast superwind.
We suggest that superwinds prevent the accretion of the ambient interstellar 
gas from the bulges and disks of their host galaxies onto the central BHs
and that in such cases the BHs are fed with the matter re-inserted by 
massive stars in
the form of numerous stellar winds and SNe explosions. The accretion rate
and the BH luminosity are then defined by the central starburst but not
by the gravity and the interstellar gas distribution in a host galaxy. 
 
The hydrodynamics of the accreted and ejected matter depend on the location
of the stellar cluster hosting a BH in the cluster mass - black hole mass -
star cluster radius parameter space. There is a surface in this 3D parameter
space which separates clusters evolving in the stationary regime from those
which cannot fulfill stationary conditions. If the mechanical luminosity
of the cluster exceeds the critical value, a hot plasma inside the cluster
is thermally unstable. The flow is highly non-stationary and presents a
complicate velocity patten with an outer stagnation point defined by 
strong radiative cooling and the inner one whose position is defined by the
mass of the BH. The hydrodynamical structure and time evolution of the
resulting flow in such cases must be calculated numerically.

Clusters whose mechanical luminosity is smaller than the critical value, 
compose stationary accretion flow in the central zones and form stationary
outflows, the star cluster winds, in the outer zones of the cluster.

We used our model to calculate the accretion rates and the accretion
luminosities of BHs at the centeres of young star forming regions.
The classic, Bondi's accretion theory shows a good agreement with our 
semi-analytic model, but only in the case of low mass clusters located well
below ($L_{SC} \le 0.1 L_{crit}$) the critical luminosity in
the $L_{SC} - M_{BH} - R_{SC}$ parameter space. Thus one has to use a 
semi-analytic approach in order to calculate the accretion rates and BH 
luminosities in the case of more energetic clusters.

In the case of extended starbursts, the BH luminosities fall well below 
the Eddington limit. However, the accretion luminosities grow rapidly for 
more compact clusters and for very compact clusters can approach the
Eddington luminosity, as shown in Figure 9, panel b.

The model, here developed, is required in order to advance our knowledge
regarding the relative contributions of supermassive BHs and central star 
bursts in composite, AGN/starburst galaxies and will be used for 
interpretation of observational properties of such objects in a further 
communication.

\acknowledgments 
We thank our anonymous referee for valuable comments and suggestions.
SS wishes to express his thanks to Isaac Shlosman for many useful 
discussions during his stay in the University of Kentucky where this 
study was designed.
This project has been supported by CONACYT - M\'exico research grants 
47534-F and 60333.


\begin{thebibliography}

\bibitem{1} Baum, S.A., O'Dea, C.P., Dallacassa, de Bruyn, A.G. \&
            Pedlar, A. 1993, ApJ, 419, 553
\bibitem{2} B\"oker, T., Laine, S.,  van der Marel, R. P., Sarzi, M.,
            Rix, H.-W., Ho, L. C., \& Shields, J. S. 2002, AJ, 123, 1389 
\bibitem{3} Bondi, H. 1952, MNRAS, 112, 195 
\bibitem{4} Cant{\'o}, J., Raga, A.C. \&  Rodr\'\i{}guez, L.F. 2000, 
            ApJ, 536, 896
\bibitem{5} Chevalier, R. A. \& Clegg, A. W., 1985, Nature, 317, 44
\bibitem{6} Collin, S. \& Zahn, J.-P. 1999, A\&A, 344, 433
\bibitem{7} C\^ot\'e P. et al. 2006, ApJS, 165, 57  
\bibitem{8} Gonz\'alez Delgado, R.M., Heckman, T., Leitherer, C.,
            Meurer, G., Krolik, J., Wilson, A.S., Kinney, A. \&
            Koratkar, A. 1998, ApJ, 505, 174
\bibitem{9} Goodman, J. 2003, MNRAS, 339, 937
\bibitem{10} Ferrarese, L.,  C\^ot\'e P., Bont\'a, E. D. et al. 2006,
             ApJ, 644, L21
\bibitem{11} Frank, J., King, A., \& Raine, D. 2002, Accretion Power in 
             Astrophysics, Cambridge, Cambridge University Press, 384p.
\bibitem{12} Heckman, T., Gonz\'alez Delgado, R.M., Leitherer, C., Meurer, G., 
             Krolik, J., Wilson, A.S.,  Koratkar, A. \&  Kinney, A. 1997,
             ApJ, 482, 114
\bibitem{13} Jim\'enez-Bail\'on, E., Santos-Lle\'o, M., Dahlem, M., Ehle, M.,
             Mas-Hesse, J.M., Guainazzi, M., Heckman, T.M. \& Weaver, K.A.
             2005, A\&A, 442, 861 
\bibitem{14} Johnson, H. E. \& Axford, W. I. 1971, ApJ, 165, 381
\bibitem{15} Leitherer, C., Schaerer, D., Goldader, J.D. et al., 
             1999, ApJS, 123, 3 
\bibitem{16} Levenson, N.A., Weaver, K.A. \& Heckman, T.M. 2001, ApJ,
             550, 230
\bibitem{17} L\'ipari, S.L. \& Terlevich, R.J. 2006, MNRAS, 368, 1001
\bibitem{18} Nulsen, P.E.J. \& Fabian, A.C. 2000, MNRAS, 311, 346
\bibitem{19} Plewa, T. 1995, MNRAS, 275, 143
\bibitem{20} Rodr\'\i{}guez Espinosa, J.M., Rudy, R.J. \& Jones, B. 1987,
             ApJ, 312, 555
\bibitem{21} Rupke, D.S., Veilleux, S. \& Sanders, D.B. 2005, ApJ, 632, 751
\bibitem{22} Sarazin, C. L. \& White, R. E. III, 1987, ApJ, 320, 32
\bibitem{23} Shlosman, I. \& Begelman, M. C. 1989, ApJ, 341, 685
\bibitem{24} Silich, S., Tenorio-Tagle, G., Rodr\'\i guez-Gonz\'alez, A. 
             2004, ApJ, 610, 226
\bibitem{25} Seth, A., Ag\"ueros, M, Lee, D. \& Basu-Zych, A. 2007,
             astro-ph/0801.0439
\bibitem{26} Tan, J.C. \& Blackman, E.G. 2005, MNRAS, 362, 983
\bibitem{27} Tenorio-Tagle, G., W\"unsch, R., Silich, S. \& Palou\v{s}, J.
             2007, ApJ, 658, 1196
\bibitem{28} Terlevich, E., Diaz, A. I. \& Terlevich, R. 1990, MNRAS,
             242, 271
\bibitem{29} Walcher, C. J., B\"oker, T., Charlot, S., Ho, L. C., Rix, H.-W.,
             Rossa, J., Shields, J. C. \& van der Marel, R. P. 2006,
             649, 692
\bibitem{30} W\"unsch, R.,  Silich, S. Palou\v{s}, J. \& Tenorio-Tagle, G.
             2007, A\&A, 471, 579
\bibitem{31} W\"unsch, R.,  Tenorio-Tagle, G., Palou\v{s}, J. \& Silich, S.
             2008, ApJ, 684 (accepted)

\end{thebibliography}
\end{document}